\documentclass[a4paper,11pt]{article}
\usepackage{pos}
\usepackage{subfigure}
\title{Stellar versus Galactic: The intensity of energetic particles at the evolving Earth and young exoplanets}
 \ShortTitle{Stellar versus Galactic cosmic rays}

\author*[a]{D. Rodgers-Lee}
\author[a]{A. A. Vidotto}
\author[b]{A. M. Taylor}
\author[c,d,e]{P. B. Rimmer}
\author[f]{T. P. Downes}

\affiliation[a]{University of Dublin, Trinity College Dublin, School of Physics,\\
  College Green, Dublin, Ireland}

\affiliation[b]{DESY,\\
D-15738 Zeuthen, Germany}

\affiliation[c]{University of Cambridge, Department of Earth Sciences,\\
 Downing St, Cambridge, United Kingdom}

\affiliation[d]{Astrophysics Group Cavendish Laboratory,\\
JJ Thomson Ave, Cambridge,  United Kingdom}

\affiliation[e]{MRC Laboratory of Molecular Biology,\\
Francis Crick Ave, Cambridge,  United Kingdom}

\affiliation[f]{Dublin City University, Centre for Astrophysics \& Relativity, \\
Glasnevin, Dublin, Ireland}

\emailAdd{drodgers@tcd.ie}
\emailAdd{aline.vidotto@tcd.ie}
\emailAdd{andrew.taylor@desy.de}
\emailAdd{turlough.downes@dcu.ie}
\emailAdd{pbr27@cam.ac.uk}

\abstract{Energetic particles may have been important for the origin of life on Earth by driving the formation of prebiotic molecules. We calculate the intensity of energetic particles, in the form of stellar and Galactic cosmic rays, that reach Earth at the time when life is thought to have begun ($\sim 3.8$Gyr ago), using a combined 1.5D stellar wind model and 1D cosmic ray model. We formulate the evolution of a stellar cosmic ray spectrum with stellar age, based on the Hillas criterion. We find that stellar cosmic ray fluxes are larger than Galactic cosmic ray fluxes up to $\sim$4\,GeV cosmic ray energies $\sim 3.8$Gyr ago. However, the effect of stellar cosmic rays may not be continuous. We apply our model to HR\,2562b, a young warm Jupiter-like planet orbiting at 20\,au from its host star where the effect of Galactic cosmic rays may be observable in its atmosphere. Even at 20\,au, stellar cosmic rays dominate over Galactic cosmic rays.}

\FullConference{37$^{\rm{th}}$ International Cosmic Ray Conference (ICRC 2021)\\
		July 12th -- 23rd, 2021\\
		Online -- Berlin, Germany}

\begin{document}
\maketitle

\section{Introduction}
Energetic particles, also known as cosmic rays, can drive prebiotic molecule formation in exoplanetary atmospheres \cite{barth_2021}. Thus, they may have been important for the origin of life on Earth which began $\sim 3.8$Gyr ago \cite{mojzsis_1996}. Cosmic rays may even have indirectly left an imprint on the helicity of DNA \cite{globus_2020}. Upcoming James Webb Space Telescope (JWST) and Ariel observations \cite{gardner_2006,tinetti_2018} will characterise exoplanetary atmospheres and search for signatures of life on other worlds. Thus, it is important to calculate the cosmic ray fluxes incident on exoplanets.

Two types of cosmic rays are important: Galactic cosmic rays from the interstellar medium (ISM) that are modulated by the stellar wind (see review by \cite{potgieter_2013} in the context of the Sun) and stellar cosmic rays (also known as stellar energetic particles) accelerated by low-mass stars like the Sun \cite{klein_2017}. Here, we consider flare-accelerated stellar cosmic rays from a solar-mass star. The combined cosmic ray flux reaching an exoplanet depends on the stellar wind properties (for example, see \cite{herbst_2020,mesquita_2021}  for Galactic cosmic ray fluxes). Young solar-type stars rotate much faster than the present-day Sun \cite{gallet_2013,gallet_2015}. The increased stellar rotation rate leads to stronger stellar magnetic field strengths \cite{vidotto_2014} and faster winds. Thus, Galactic cosmic ray fluxes at Earth's orbit should become more suppressed by the magnetised stellar wind for decreasing stellar ages. In contrast, stellar cosmic rays become more important with decreasing stellar age \cite{rab_2017, rodgers-lee_2017}.

Here, we quantify the contribution from stellar and Galactic cosmic rays at Earth's orbit at approximately the time when life began on Earth. We also present the cosmic ray intensities reaching HR\,2562b, a young (600\,Myr) Jupiter-like planet orbiting at a large distance from its host star \cite{konopacky_2016}. We use a combined 1.5D stellar wind model \cite{johnstone_2015a,carolan_2019} and a 1D diffusion-advection cosmic ray transport equation \cite{parker_1965} to model the propagation of stellar and Galactic cosmic rays through the stellar wind for different stellar ages. The model is presented in \cite{rodgers-lee_2020b,rodgers-lee_2021}. The wind properties for the Sun-like star are evolved with rotation rate, the scaling laws that we use are also discussed in \cite{rodgers-lee_2020b}.
\begin{figure}
	\centering
        \includegraphics[width=0.7\textwidth]{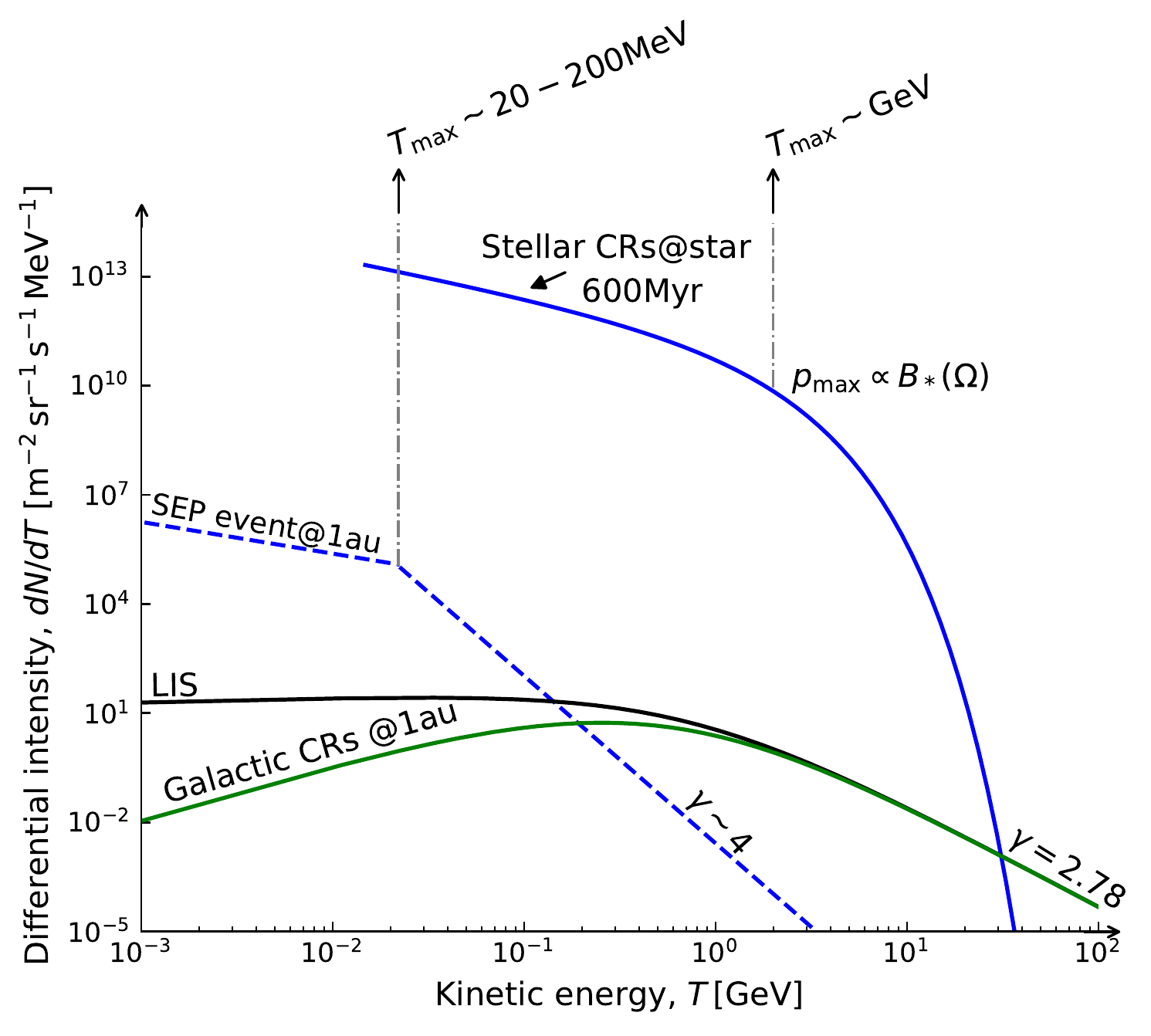}
       	\centering
  \caption{Differential intensity of Galactic and stellar cosmic rays as a function of cosmic ray kinetic energy. The LIS is shown by the black line and the green line represents present-day Galactic cosmic ray intensities at Earth. The stellar cosmic ray spectrum that we motivate for $t_\star=600$\,Myr close to the stellar surface is shown by the solid blue line. The dashed blue line is a time-averaged typical solar energetic particle spectrum observed at Earth\label{fig:sketch}. Figure from \cite{rodgers-lee_2021}.} 
\end{figure}
\section{Cosmic ray spectra}
We adopt a fit from \cite{vos_2015} to the local interstellar spectrum (LIS) as a fixed outer boundary condition for the Galactic cosmic rays for all stellar ages that we investigate. The size of the astrosphere (stellar analogue of Sun’s heliosphere) increases with decreasing stellar age due to the increasing stellar wind ram pressure (the ISM ram pressure is held constant throughout). The LIS is shown by the black line in Fig.\,\ref{fig:sketch}. Galactic cosmic ray intensities representative of present-day values observed at Earth are shown by the green line in Fig.\,\ref{fig:sketch}. 

For the stellar cosmic ray spectrum, we increase the maximum stellar cosmic ray energy with decreasing stellar age using the Hillas criterion \cite{hillas_1984}. We assume 10\% of the stellar wind kinetic power is available to accelerate stellar cosmic rays. The stellar wind kinetic power increases with decreasing stellar age due to an increase in the stellar wind velocities. We adopt a power-law spectral index of $-2$, representative of diffusive shock acceleration and compatible with magnetic reconnection as the acceleration process close to the stellar surface. Further details for the stellar cosmic ray spectrum are given in \cite{rodgers-lee_2021}.

\section{Energetic particle fluxes at the time when life began on Earth}
The cosmic ray differential intensities at a stellar age of $t_\star=1$\,Gyr are shown in Fig.\ref{fig:1gyr}. The blue and green shaded regions correspond to the stellar and Galactic cosmic ray fluxes found in the habitable zone, respectively. The Galactic cosmic ray fluxes are much smaller than the present-day values observed at Earth (see Fig.\,\ref{fig:sketch}) due to the strong magnetic field and faster stellar wind at $t_\star=1$\,Gyr. The solid black line is the LIS fit and the grey dash-dotted line is the pion threshold energy, 290\,MeV. Cosmic rays with energies greater than the pion threshold produce particle showers in Earth-like atmospheres and the secondary particles can reach the surface of the planet. At this stellar age, stellar cosmic ray intensities are larger than Galactic cosmic intensities up to $\sim$4\,GeV energies. Thus, stellar cosmic rays were likely more important than Galactic cosmic rays when life began on Earth. It is important to note that the effect of the stellar cosmic rays may not have been continuous (see \cite{rodgers-lee_2021} for a discussion of this point).

\section{Energetic particle fluxes for HR\,2562b}
\begin{figure*}%
	\centering
    \subfigure[]{%
        \includegraphics[width=0.5\textwidth]{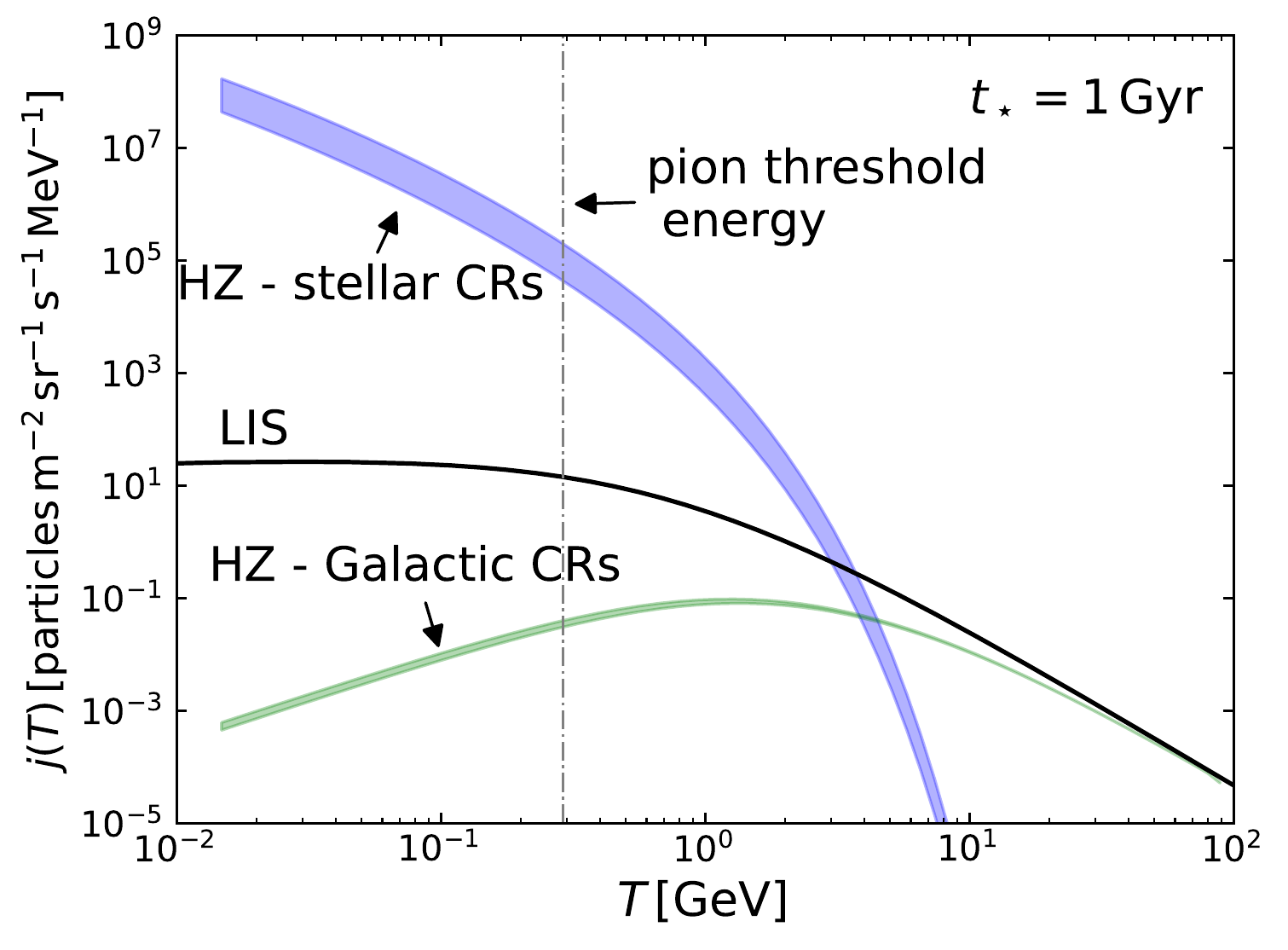}
       	\centering
\label{fig:1gyr}}%
~
	\centering
    \subfigure[]{%
        \includegraphics[width=0.5\textwidth]{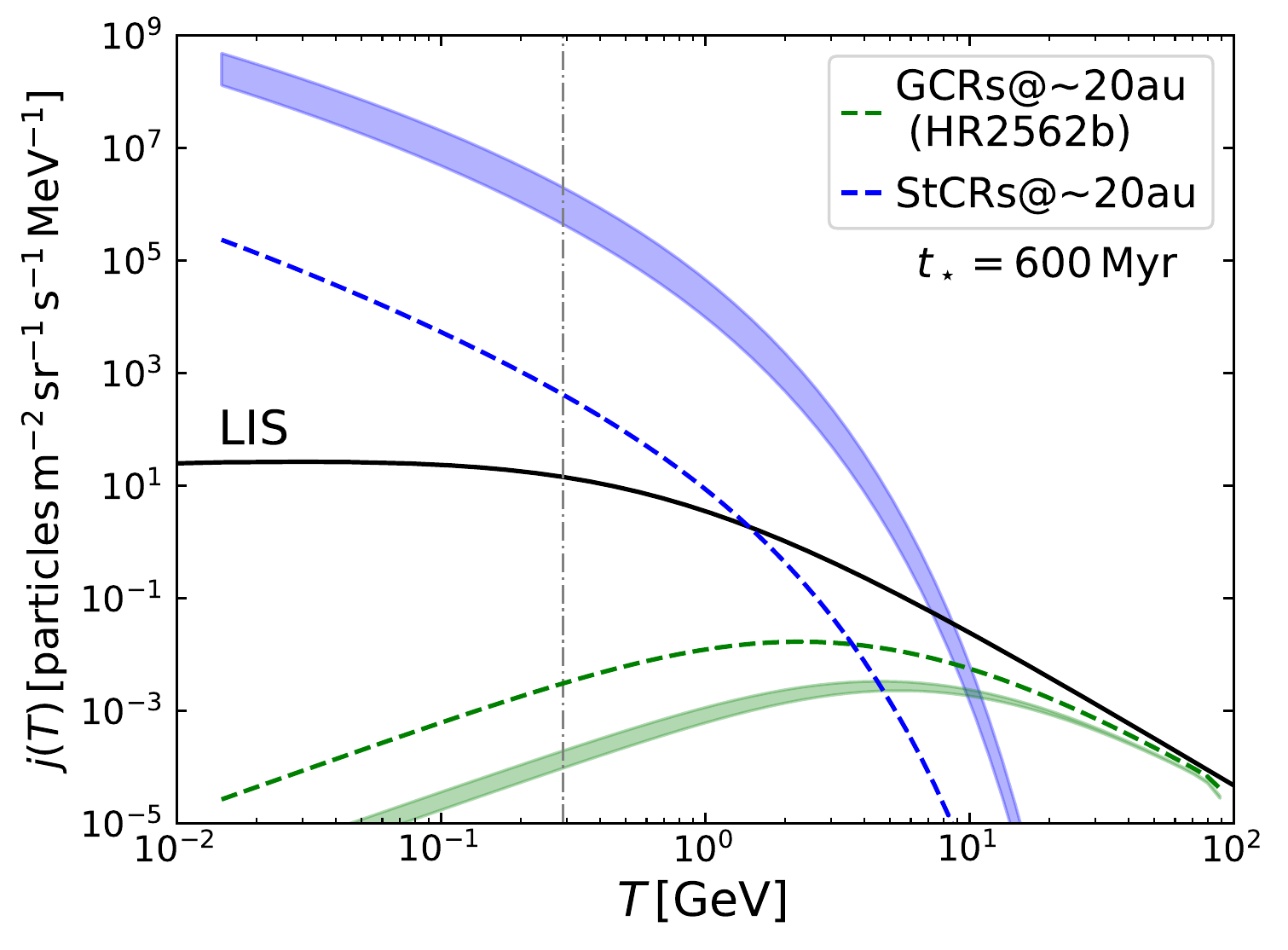}
       	\centering
\label{fig:600myr}}%
    \caption{The cosmic ray differential intensities for (a) $t_\star=1\,$Gyr and (b) $t_\star=600$\,Myr. The blue and green shaded regions represent the values for stellar and Galactic cosmic rays, respectively, in the habitable zone. In (b) the dashed green and blue lines represent the same quantities at the orbital distance of HR\,2562b. The pion threshold energy is denoted by the grey dash-dotted line. Figure adapted from \cite{rodgers-lee_2021}.} 
    \label{fig:ages}%
\end{figure*}

The cosmic ray differential intensities at $t_\star=600$\,Myr are shown in Fig.\ref{fig:600myr}, similar to Fig.\ref{fig:1gyr}. The stellar and Galactic cosmic ray intensities at the orbital distance of HR\,2562b ($\sim$20\,au) are shown by the dashed blue and green lines, respectively. Again, stellar cosmic ray fluxes may still be important, even at this distance but are more likely to be transient in time. By constraining abundances of molecules such as $\mathrm{H_3O^+}$ \cite{helling_2019} in an atmosphere like HR\,2562b with JWST it may be possible to constrain the incident cosmic ray flux. The Galactic cosmic ray fluxes at different heights in an atmosphere representative of HR\,2562b are shown in \cite{rodgers-lee_2020b}.

\section{Conclusions}
Here we presented the cosmic ray intensities for different stellar ages, published in \cite{rodgers-lee_2020b,rodgers-lee_2021}. At $t_\star=1\,$Gyr, approximately the time when life is thought to have begun on Earth, we found that stellar cosmic ray fluxes are larger than Galactic cosmic ray fluxes up to $\sim4\,$GeV cosmic ray energies. However, the stellar cosmic ray fluxes may not be continuous in time at GeV energies. We also showed that stellar cosmic rays dominate over Galactic cosmic rays for HR\,2562b, a young Jupiter-like planet orbiting a solar-type star. Warm Jupiters, such as HR\,2562b, represent good candidates for future JWST observations to detect the chemical signature of cosmic rays in an exoplanetary atmosphere.

\section*{Acknowledgements}
DRL and AAV acknowledge funding from the European Research Council (ERC) under the European Union’s Horizon 2020 research and innovation programme (grant agreement No 817540, ASTROFLOW). P. B. R. thanks the Simons Foundation for support under SCOL awards 59963. DRL wishes to acknowledge the DJEI/DES/SFI/HEA Irish Centre for High-End Computing (ICHEC) for the provision of computational facilities and support.

\newcommand\aj{AJ} 
\newcommand\actaa{AcA} 
\newcommand\araa{ARA\&A} 
\newcommand\apj{ApJ} 
\newcommand\apjl{ApJ} 
\newcommand\apjs{ApJS} 
\newcommand\aap{A\&A} 
\newcommand\aapr{A\&A~Rev.} 
\newcommand\aaps{A\&AS} 
\newcommand\mnras{MNRAS} 
\newcommand\pasa{PASA} 
\newcommand\pasp{PASP} 
\newcommand\pasj{PASJ} 
\newcommand\solphys{Sol.~Phys.} 
\newcommand\nat{Nature} 
\newcommand\bain{Bulletin of the Astronomical Institutes of the Netherlands}
\newcommand\memsai{Mem. Societa Astronomica Italiana}
\newcommand\apss{Ap\&SS} 
\newcommand\qjras{QJRAS} 
\newcommand\pof{Physics of Fluids}
\newcommand\grl{Geophysical Research Letters}
\newcommand\planss{Planetary and Space Science}
\newcommand\ssr{Space Science Reviews}
\newcommand\astrobiology{Astrobiology}
\newcommand\icarus{Icarus}
\newcommand\jgr{Journal of Geophysical Research}

\bibliographystyle{JHEP}
\bibliography{donnabib}

%
%
%

\end{document}